\documentclass[prl,aps,showpacs,twocolumn,unsortedaddress]{revtex4}
\usepackage{graphics, bm}
\usepackage{psfrag}
\usepackage{amsmath}
\usepackage{amssymb}
\usepackage{epsfig}
\newcommand{\beq}    {\begin{equation}}
\newcommand{\enq}    {\end{equation}}
\newcommand{\ceq}[1] {(\ref{#1})}
\newcommand{\rr}     {{\bf r}}

\DeclareMathOperator{\sgn}{sgn}
\newcommand{\df}     {\equiv}
\newcommand{\nav}    {\langle n\rangle}
\begin{document}

\title{Ground-state of graphene in the presence of random charged impurities.}

\author{Enrico Rossi, S. Das Sarma}
\affiliation{Condensed Matter Theory Center, Department of Physics,\\
             University of Maryland, College Park, MD 20742-4111, USA}
\date{\today}

   
\begin{abstract}
We calculate the carrier density dependent ground state properties of
graphene in the presence of random charged impurities in the substrate
taking into account disorder and interaction effects
non-perturbatively on an equal footing in a self-consistent
theoretical formalism.  We provide detailed quantitative results on
the dependence of the disorder-induced spatially inhomogeneous
two-dimensional carrier density distribution on the external gate
bias, the impurity density, and the impurity location.  We find that
the interplay between disorder and interaction is strong, particularly
at lower impurity densities.  We show that for the currently available
typical graphene samples, inhomogeneity dominates graphene physics at
low ($\lesssim 10^{12}$~cm$^{-2}$) carrier density with the density
fluctuations becoming larger than the average density.
\end{abstract}


\maketitle


The recent experimental realization \cite{experiments} of single-layer
graphene sheets has spurred an enormous amount of activity in studying
the electronic properties of 2D chiral Dirac fermions in the context
of solid state materials physics.  While much of this interest is
fundamental, a substantial part of it also derives from the
technological prospect of graphene being used as a novel transistor
material.  To understand current experiments and be able to design
future graphene-based electronic devices it is essential to know the
properties, origin and effects of extrinsic disorder in graphene.  The
low energy electronic states of graphene are described by a massless
Dirac equation.  In clean isolated graphene (the so-called intrinsic
graphene) the Fermi energy lies exactly at the Dirac point (i.e. the
charge neutrality point) where the linear chiral electron and hole
bands cross each other.  Several works \cite{intrinsic_refs}
calculated the graphene conductivity assuming the graphene Fermi
energy to be exactly at the Dirac point throughout the graphene layer.
These works found the Dirac point conductivity to be either 0, $\infty$
or, in the limit of vanishing disorder, equal to the universal value
$\sigma_D\equiv 4e^2/\pi\hbar$. In current experiments
however the measured conductivity at the Dirac point \cite{tan} is
finite and much bigger (by a factor of 2-20) than the universal theoretical
prediction $\sigma_D$ and varies strongly from sample to sample.  The
discrepancy can be resolved if we consider that disorder in addition
to represent the main source of scattering has an other important
effect: it locally shifts the Dirac point removing, at zero gate
voltage, the Fermi energy from the charge neutrality point
\cite{Geim:2006}.  This leads immediately to a disorder-induced
inhomogeneous density landscape with electron-hole puddles.  Such
puddles have been proposed theoretically \cite{hwang} and observed
experimentally \cite{yacoby, brar}.  Experiments, by themselves, are
unable to directly identify the cause of the carrier density
inhomogeneities.  Two kinds of disorder have been proposed in graphene
to have this effect: ripples \cite{ripples} and random charge
impurities \cite{hwang}.  Transport theories
\cite{hwang,nomura,shaffique,ando} based on the presence of charge
impurities have been successful in explaining the experimental results
\cite{tan}.  But whether the puddles arise from the random charged
impurities or from some other mechanism \cite{ripples} has remained an
open question.  We provide in this Letter the first realistic
theoretical description of the electron-hole puddles in graphene
assuming the random charged impurity disorder to be the underlying
mechanism.  Our theoretical results are in excellent qualitative
agreement with the existing experimental data \cite{yacoby, brar}.  A
quantitative comparison between our results and future experiments
with higher quantitative accuracy would enable a definitive
understanding of the nature of the disorder in graphene.

At low energies the quasiparticles in graphene can be described by a
massless Dirac-fermion, MDF, model with an ultraviolet cutoff wave
vector $k_c$. We set $k_c = 1/a_0$, where $a_0$ is the graphene
lattice constant, $a_0=0.246\;{\rm nm}$, corresponding to an energy
cut-off $E_c\approx 3\;{\rm eV}$, and measure the energies from the
Dirac point.  To find the ground-state carrier density $n$ we use the
Thomas-Fermi-Dirac (TF) theory. In contrast with the standard TF
theory we retain the exchange potential non perturbatively through its
local density approximation so that the energy functional $E[n]$ reads:
\begin{align}
 \!E[n]\!\! &= \!\hbar v_F\!\!\left[\!\frac{2\sqrt{\pi}}{3}\!\!\!\int\!\!\!d^2 r\sgn(n)|n|^{3/2}\!\!  
        + \!\frac{r_s}{2}\!\!\int\!\!\! d^2 r\!\!\!
          \int\!\!\!d^2 r'\! \frac{n(\rr)n(\rr'\!)}{|\rr - \rr'|}\right.\nonumber \\ 
        & \;\;+\!\left.\frac{E_{xc}[n]}{\hbar v_F}
          \!+\!r_s\!\!\int\!\! d^2 r V_D(\rr)n(\rr)\! -\! 
          \frac{\mu}{\hbar v_F}\!\int\!\! d^2 r n(\rr)\right]
 \label{eq:en}
\end{align}
where $v_F = 10^6\;{\rm m/s}$ is the Fermi velocity, $r_s\df
e^2/(\hbar v_F\epsilon)$ is the coupling constant with $\epsilon$ the
effective background dielectric constant, $E_{xc}[n]$ is the exchange
energy, $V_D$ is the disorder potential and $\mu$ is the chemical
potential.  The first two terms in \ceq{eq:en} are the kinetic energy
and the Hartree part of the Coulomb interaction, respectively.  For
graphene on ${\rm SiO_2}$ substrate $\epsilon = 2.5$ and then
$r_s=0.8$.  
By differentiating $E[n]$ with respect to $n$ we find:
\begin{equation}
 \frac{\delta E}{\delta n}\!=\!\hbar v_F\!\!\left[\!\sgn(n)\sqrt{\!\pi |n|}\!+\!
                             \frac{r_s}{2}\!\!\int\!\!\frac{n(\rr')d^2\!r'}{|\rr\!-\!\rr'|}\!+\!r_s\!V_D\! 
                             \right]
                             \!+\!\Sigma(n)\!-\!\mu
 \label{eq:de_dn}
\end{equation}
where $\Sigma(n)$ is the Hartree-Fock self-energy
\cite{barlas,hwang_hu} evaluated at the Fermi wave vector
$k_F = \sgn(n)\sqrt{\pi |n}|$:
\begin{equation}
 \frac{\Sigma(n)}{\hbar v_F}
               \!=\!\sqrt{\pi |n|}\sgn(n) r_s\!\left[
                \frac{1}{4}\ln\!\frac{4 k_c}{\sqrt{\pi |n|}}
               \!-\!\!\left(\!\frac{2C\!+\!1}{2\pi}\!+\!\frac{1}{8}\!\right) 
               \right]
 \label{eq:sigma}
\end{equation}
where $C\approx 0.916$.

We assume $V_D$ to be the 2D Coulomb potential in the graphene plane
generated by a random 2D distribution, $C(\rr)$, of impurity charges
placed at a distance $d$ from the graphene layer.  Denoting by angular
brackets the average over disorder realizations we assume:
\begin{equation}
 \langle C(\rr)\rangle = 0; \hspace{0.5cm}
 \langle C(\rr_1) C(\rr_2)\rangle = n_{imp}\delta(\rr_2 - \rr_1);
 \label{eq:C_stat}
\end{equation}
where $n_{imp}$ is the 2D impurity density.  A non zero value of
$\langle C(\rr)\rangle$ can be taken into account by a shift of $\mu$,
i.e. of the gate voltage.  $n_{imp}$ and $d$ should be taken as
effective parameters characterizing the impurity distribution in a
minimal two parameter model. In current graphene samples obtained
through mechanical exfoliation possible sources of charge impurities
are most likely ions in the substrate that drift close to the surface,
charges trapped between the graphene layer and the substrate and free
charges that stick to the top surface of the graphene layer.  This
picture is consistent with the vast literature on disorder in Si
MOSFET and has recently been indirectly confirmed by experiments
on suspended graphene \cite{suspended}.  The values of $n_{imp}$
extracted from transport measurements, and used in this work, are
indeed of the same order of magnitude, $[10^{11}-10^{12}]$~cm$^{-2}$,
as the ones used to describe quantitatively disorder effects of MOSFET
devices on SiO$_2$.  Combining Eq.\ceq{eq:de_dn}-\ceq{eq:C_stat} we find
the ground state carrier density by solving the equation $\delta
E/\delta n = 0$ using the steepest descent method.  Our calculations
are done for a finite square lattice of size $L\times L$.  All the
results presented in this Letter are obtained for $L = 200$~nm and are
found to be independent of system size for $L\gtrsim100$~nm.  For the
discretization in real space we use a 1~nm step.

\begin{figure}[htb]
 \begin{center}
  \includegraphics[width=8.5cm]{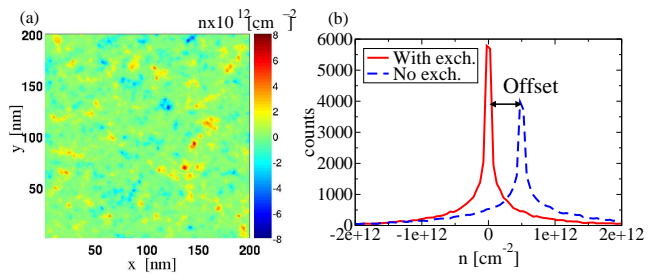}
  \caption{
           (Color online). 
	   Results at the Dirac point
	   for a disorder realization assuming 
	   $n_{imp}=10^{12}\;{\rm cm^{-2}}$, $d=1$~nm,
	   $\epsilon = 2.5$.
	   (a) Color plot of $n(\rr)$ including exchange.
	   (b) Density distribution for $n(\rr)$  shown in (a).
	       For clarity the result without
	       exchange has been offset along the $x$-axis.
          } 
  \label{fig:1}
 \end{center}
\end{figure}

For a given disorder realization, for $\mu=0$, a typical result,
including exchange, for $n(\rr)$ is shown in Fig.~\ref{fig:1}~(a).
For $n>0$ ($n<0$) we have particles (holes).  The result without
exchange is characterized by larger density fluctuations.  This is
clear from Fig.~\ref{fig:1}~(b) which shows that the density
distribution is more strongly peaked around $n=0$ when exchange is
taken into account.  The result of Fig.~\ref{fig:1}~(b) is
counter-intuitive because exchange suppresses density inhomogeneity
instead of enhancing it as in parabolic-band inhomogeneous electron
liquids.
A complementary DFT-LDA calculation, using single disorder
realizations with few impurities, has also found similar results
\cite{polini}. Contrary to our work in \cite{polini} the correlation
contributions have been taken into account.  Given the numerical
complexity of the DFT-LDA approach in \cite{polini} only small samples
were considered and disorder averaged results, that would permit a
close quantitative comparison, were not presented. The results for
single disorder realizations are qualitatively similar to ours,
showing that correlation terms have only a minor quantitative effect.
The reason is that in graphene to very good approximation the correlation term
scales with $n$ in the same way as exchange \cite{barlas, polini} but
with opposite sign and therefore its effect is to simply reduce the
exchange strength. 

\begin{figure}[htb]
 \begin{center}
  \includegraphics[width=8.5cm]{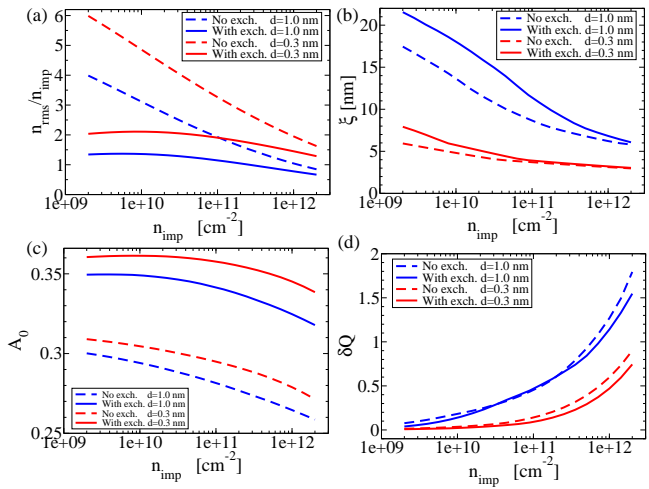}
  \caption{(Color online).
           Results as a function of $n_{imp}$ at the Dirac point
           for $d=1$~nm (blue lines) and $d=0.3$~nm (red lines).
	   $\epsilon=2.5$.
           (a) $n_{rms}$; (b) $\xi$ ;
	   (c) $A_0$; (d) $\delta Q$. 
          } 
  \label{fig:3}
 \end{center}
\end{figure}

The results of Fig.~\ref{fig:1} are visually very similar to the ones
observed in experiments \cite{brar,yacoby}, but a quantitative
comparison can only be achieved by calculating the disorder-averaged
statistical properties.  For a given quantity $X$ we therefore
calculate its disorder averaged value, $\langle X\rangle$, and spatial
correlation function $\langle(\delta X(\rr))^2\rangle=\langle
(X(\rr)-\langle X\rangle) (X(0) -\langle X\rangle)\rangle$.  From
these results we extract the rms of the fluctuations,
$X_{rms}\df\sqrt{\langle(\delta X(0))^2\rangle}$, and their typical
correlation length $\xi_X\df FWHM$ of $\langle(\delta
X(\rr))^2\rangle$.  
At the neutrality point $\xi\df\xi_n$ can be
loosely taken as a measure of the electron-hole puddle size.  
%
%
In Fig.~\ref{fig:3} we present the disordered averaged results at the
Dirac point as a function of $n_{imp}$. We see that exchange
suppresses the amplitude of the density fluctuations and increases
their correlation length and that its effect becomes increasingly
important as the impurity density decreases; for the lowest $n_{imp}$
the value of $n_{rms}$ including exchange is 3 times smaller than the
value obtained without exchange, Fig.~\ref{fig:3}~(a).  In addition we
see that the scaling of $n_{rms}$ with $n_{imp}$ is very different
with and without exchange.  From Fig.~\ref{fig:3}~(b) we see that 
as $n_{imp}$ decreases $\xi$ increases
very slowly, especially for low values of $d$, a  result that
underlines the
importance of non-linear screening terms.  Adapted to a 2D
distribution of charges the approach used in Ref.~\cite{shklovskii}
for the scaling of $\xi$ on $n_{imp}$ gives $\xi\approx
1/(r_s^2\sqrt{n_{imp}})$.  For $r_s=0.8$ and
$n_{imp}=2.\;10^9$~cm$^{-2}$ we would then expect $\xi\approx 350$~nm,
a value an order of magnitude larger than the value shown in
Fig.~\ref{fig:3}~(b).  The reason for this discrepancy is that for
small values of $n_{imp}$ the carrier distribution is not
characterized by smooth long-range fluctuations but rather by wide
regions of very small carrier density ($\approx 0$) interspersed with
small electron-hole puddles with the typical size $\xi$ shown in
Fig.~\ref{fig:3}~(b).  This picture is confirmed in
Fig.~\ref{fig:3}~(c) where the disorder averaged area fraction, $A_0$,
over which $|n(\rr)-\langle n\rangle|<n_{rms}/10$ is plotted as a
function of $n_{imp}$.  We see that as $n_{imp}$ decreases $A_0$
increases reaching more than 1/3 at the lowest impurity densities.
The fraction of area over which $|n(\rr)-\langle n\rangle|$ is less
than $1/5$ of $n_{rms}$ surpasses 50\% for $n_{imp}\lesssim
10^{10}$~cm$^{-2}$.  Thus, much of the 2D landscape in this situation
has very low ($\ll n_{\rm imp})$ carrier density with a few random electron-hole
puddles.  In Fig.~\ref{fig:3}~(d) the dependence of the excess charge
$\delta Q\df n_{rms}\pi\xi^2$ on $n_{imp}$ is shown.  At high impurity
densities ($\gtrsim 10^{12}$~cm$^{-2}$) and values of $d\gtrsim 1$~nm,
$\delta Q$ can be approximately identified with the average number of
carriers per puddle, however the above discussion and the results for
$A_0$ allow us to recognize that $\delta Q$, for small $n_{imp}$, {\it
is not} the typical number of carriers per puddle.  The reason is that
for small $n_{imp}$ (and/or $d$), because of the large fraction of
area over which is $|n(\rr)-\langle n\rangle|\ll n_{rms}$, $n_{rms}$
is much smaller than the typical carrier density in a electron-hole
puddle of size $\xi$.  At low $n_{imp}$ $\delta Q$ grossly
underestimates the number of carriers in a typical puddle of size $\xi$.

We are now in a position to discuss the validity of our TF
approach. The use of the TF theory is justified when the condition
$|\nabla n(\rr)|/[n(\rr)k_F(\rr)]\ll 1$ is satisfied.  If we estimate
$|\nabla n(\rr)|\approx n_{rms}/\xi$, the above inequality implies
$\sqrt{\pi n_{rms}}\xi\gg 1$, i.e. $\delta Q \gg 1$.  However, for
small $n_{imp}$ and $d$, $n_{rms}$ greatly underestimates $|n|$ in the
regions where it is inhomogeneous, i.e. in the electron-hole puddles
of size $\xi$. We find that at low $n_{imp}$, $|n|$ in these
electron-hole puddles is a factor of 10 or more higher than $n_{rms}$.
This can already be seen for relatively high values of $n_{imp}$ and
$d$: from Fig.~\ref{fig:1}~(a) we see that $|n|$ inside the
electron-hole puddles takes values as high as $8.\; 10^{12}$~cm$^{-2}$
whereas the corresponding value of $n_{rms}$ is only $8.\;
10^{11}$~cm$^{-2}$, Fig.~\ref{fig:3}~(a).  Even in the limiting case
of an isolated impurity with $d=0$ the density profile obtained using
the TF approach \cite{katsnelson:2006} is very similar to the one
obtained starting from the Dirac equation and treating the interaction
via RG \cite{bisaws:2007}.  The additional $\delta(\rr)$ for $n(\rr)$
found in \cite{bisaws:2007} (and \cite{shytov:2007}) in real graphene,
in which the MDF model applies only at low energies, is regularized by
max$[d, a_0]$, \cite{Fogler:2007}.  Our results are therefore
quantitatively accurate.

From the theoretical analysis \cite{hwang, shaffique, nomura, ando} of
experimental transport results \cite{tan} one obtains for typical
graphene samples on SiO$_2$, $d=1$~nm and $n\approx3.\;10^{11}$~cm~$^{-2}$.  
For these values, from Fig.~\ref{fig:3}~(a)
and (b) we see that $n_{rms}=3.\;10^{11}$~cm~$^{-2}$ and $\xi=9$~nm.
The value of $n_{rms}$ is in very good agreement with the recent STM
\cite{brar} and SET \cite{yacoby} results. The value of $\xi$ is also
in very good agreement with the STM results and consistent with the
results of Ref.~\cite{yacoby} that, given the lower SET spatial resolution ($\gtrsim
150$~nm), could only provide for $\xi$ an upper bound of 30~nm.

\begin{figure}[!!!!!!!!!!!!!!!!h!!!!!!!!!!t]
 \begin{center}
  \includegraphics[width=8.5cm]{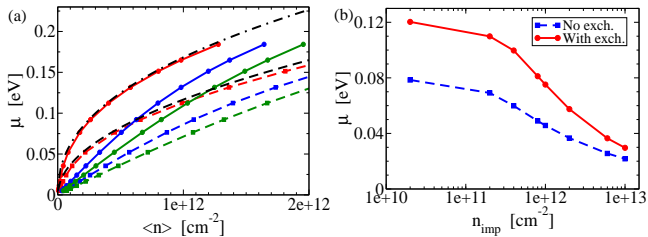}
  \caption{(Color online).
           Solid (dashed) lines: results with (without) exchange.
	   $\epsilon=2.5$.
           (a) $\mu$ vs. $\nav$ for 
               $n_{imp}=2.10^{10}\;{\rm cm^{-2}}$, red lines;
	       $n_{imp}=10^{12}\;{\rm cm^{-2}}$,   blue lines;
	       $n_{imp}=2.10^{12}\;{\rm cm^{-2}}$, green lines.
	       The black dot-dash (dash) line shows $\mu(n)$ 
	       for clean graphene with (without) exchange.
	   (b) $\mu$ vs. $n_{imp}$ for $\nav=5\;10^{11}\;{\rm cm^{-2}}$.
          } 
  \label{fig:4}
 \end{center}
\end{figure}

\begin{figure}[htb]
 \begin{center}
  \includegraphics[width=8.5cm]{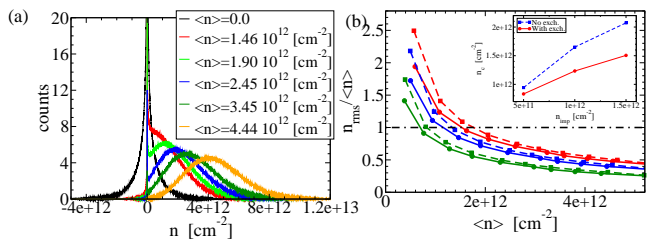}
  \caption{(Color online).
           Results away from the Dirac point assuming ${\rm SiO_2}$ substrate.
	   (a) Density distribution averaged over disorder
	       for different values of $V_g$
	       for $d=1$~nm and $n_{imp}=10^{12}\;{\rm cm^{-2}}$.
	   (b) $n_{rms}/\langle n\rangle$ vs. $\langle n\rangle$ for $d=1$~nm and:
               $n_{imp} = 1.5\times 10^{12}\;{\rm cm^{-2}}$, red lines;
	       $n_{imp} = 10^{12}\;{\rm cm^{-2}}$, blue lines;
	       $n_{imp} = 5\times 10^{11}\;{\rm cm^{-2}}$, green lines.
	       Inset: $n_c$ vs. $n_{imp}$ for $d=1$~nm.
	       Solid (dashed) lines: results with (without) exchange.
          } 
  \label{fig:da_n}
 \end{center}
\end{figure}

At a finite gate voltage, $V_g$, the average carrier density $\langle
n\rangle=C_g V_g/e$ is induced, where $C_g$ is the gate capacitance.
In our calculations we indirectly fix $\langle n\rangle$ by varying
the chemical potential $\mu$.  The relation between $\mu$ and $\langle
n\rangle$ is shown in Fig.~\ref{fig:4}~(a).  Contrary to ordinary
parabolic band fermionic systems the relation between $\mu$ and $\nav$
strongly depends on disorder even when exchange is neglected.  This is
also shown in Fig.~\ref{fig:4}~(b) in which $\mu$ is plotted as a
function of $n_{imp}$ for a fixed value of $\nav$.  The dependence of
$\mu(\nav)$ on $n_{imp}$ even without exchange is due to the fact that
in graphene the kinetic energy does not scale linearly with $n$.  From
Fig.~\ref{fig:4}~(a) we see that only for $n_{imp}\lesssim
10^{10}$~cm$^{-2}$ $\mu(\nav)$ follows the equation valid for clean
graphene.  We also notice that $\mu(\nav)$ is strongly affected by
exchange.  The results of Fig.~\ref{fig:4} demonstrate the interplay
of disorder and interaction in graphene and show how the dependence of
$\mu$ on $\nav$, and in particular the average compressibility $1/(n^2
\partial\mu/\partial n)$, can be used to probe the strength of
disorder and many-body effects.
In Fig.~\ref{fig:da_n}~(a) the disorder averaged density distribution
obtained including exchange is plotted for different values of
$\langle n\rangle$, i.e. of $V_g$.  For $V_g=0$ the distribution has a
strong peak ($\approx$ 20 times the maximum of the y-scale of
Fig.~\ref{fig:da_n}~(a)) at $n=0$.  As $V_g$ increases the $n=0$ peak
survives and a broad peak at finite $n$ develops. For large enough
$V_g$ the $n=0$ peak disappears and the density distribution is
characterized only by the broad peak centered at $n=\langle n\rangle$.
The results without exchange are qualitatively similar.  The double
peak structure for finite $V_g$ provides direct evidence for the
existence of puddles over a finite voltage range.  High values of
$V_g$ remove one kind of puddles and increase the amplitude of the
density fluctuations reflected in an increase of $n_{rms}$.  On the
other hand the ratio $n_{rms}/\langle n\rangle$ decreases
monotonically as a function of $\langle n\rangle$ as can be seen in
Fig.~\ref{fig:da_n}~(b).  We can define a characteristic density,
$n_c$, as the value of $\langle n\rangle$ for which $n_{rms}=\langle
n\rangle$ with $\Delta V_g = e n_c/C_g$ loosely measuring the width in
gate voltage over which the transport properties of graphene are
dominated by the density fluctuations around the Dirac point.  The
inset of Fig.~\ref{fig:da_n}~(b) shows $n_c$ as a function of
$n_{imp}$ for $d=1$~nm. We can see that  in current samples $n_{rms}\gtrsim \nav$ for carrier
densities as high as $\nav\approx 10^{12}$~cm$^{-2}$.
The particular dependence of the carrier
density distribution and $n_{rms}$ on $V_g$ are unique to
inhomogeneities created by charged impurities and is a prediction that
should be easy to verify experimentally.

We conclude by summarizing our key qualitative findings: (1) both
disorder and many-body effects become quantitatively very important on
the chemical potential close to the Dirac point; (2) many body effects
are more important at lower values of $n_{imp}$; (3) for low $n_{imp}$
the ground state near the Dirac point is characterized by small
puddles and large regions of almost zero $(|n|\ll n_{imp})$ carrier
density; (4) in current samples $n_{rms}\gtrsim \nav$ for carrier
densities as high as $\nav\approx 10^{12}$~cm$^{-2}$; (5) the number
of carriers per puddle is $\sim$1-5 at low carrier densities; (6) our
theory agrees well with the existing data \cite{yacoby, brar} but more
experiments will be required to test our quantitative predictions.

We thank S. Adam, M. Fuhrer, E.~H. Hwang, and especially
A.~H. MacDonald for discussions.  The numerical calculations have been
performed on the University of Maryland High Performance Computing
Cluster (HPCC).  This work is supported by NSF-NRI-SWAN and US-ONR


%

%
%
\end{document}